\documentclass[a4paper]{jpconf}

\usepackage{citesort}

\usepackage{graphicx}
\usepackage{amsmath} 

\begin{document}
\title{Richardson's solutions in the real- and complex-energy spectrum}

\author{Rodolfo M. Id Betan}

\address{Instituto de F\'isica Rosario (CONICET-UNR), Bv. 27 de Febrero 210 bis, S2000EZP Rosario. Argentina}
\address{Facultad de Ciencias Exactas, Ingenier\'ia y Agrimensura (UNR), Av. Pellegrini 250, S2000BTP Rosario. Argentina}
\address{Instituto de Estudios Nucleares y Radiaciones Ionizantes (UNR), Riobamba y Berutti, S2000EKA Rosario. Argentina}

\ead{idbetan@ifir-conicet.gov.ar}

\begin{abstract}
The  constant pairing Hamiltonian holds exact solutions worked out by Richardson in the early Sixties.  This exact solution of the pairing Hamiltonian regained interest at the end of the Nineties. The discret complex-energy states had been included in the Richardson's solutions in Ref. \cite{2003Hasegawa}. In this contribution we reformulate the problem of determining the exact eigenenergies of the pairing Hamiltonian when the continuum is included through the single particle level density. The solutions with discret complex-energy states is recovered by analytic continuation of the equations to the complex energy plane. This formulation may be applied to loosely bound system where the correlations with the continuum-spectrum of energy is really important. Some details are given to show how the many-body eigenenergy emerges as sum of the pair-energies.
\end{abstract}

\section{Introduction}
It is an usual practice, in many-body calculation, starts from a single-particle representation. Probably, the best way to define this basis, is from the eigenfunctions of the single-particle Hamiltonian obtained from the many-body mean-fiel  \cite{2007Suhonen}. For loosely-bound systems, it is expected that the eigenenergies of the single-particle Hamiltonian, will lay very close to the continuum threslhold. This means that the Fermi level will be also very close to continuum or even it may lies in the continuum itself. For these kind of system, correlations with continuum spectrum of energy becomes very important and so they have to be taken into account explicitly.

The residual interaction coming from the mean-field approximation contains all possible correlations between the different configurations. This implies that the many-body basis will, rapidly increase with the number of particles in the system. Besides, the need to include the continuum throws the dimension to huge numbers. The pairing interaction gives a good approximation to the residual interaction \cite{1964Lane,2003Dean}. It reeplaces the matrix diagonalization procedure by a roots finding procedure from a small set of algebraic non-linear equations.

The independent state pairing interaction admits exact solution that was worked out in 1963 by Richardson \cite{1963Richardson}; only a few years after that Bardeen, Cooper and Shrieffer \cite{1957Bardeen} explained microscopically the superconductivity as the pairing of electrons in the neighborhood of the Fermi level.  

The exact treatment of the pairing correlations with continuum representation in many-body systems, allows to assess the importance of the continuum-continuum correlations in systems with many particles. This is achieved by solving the Richardson equations with resonant and non-resonant configurations \cite{2012npaIdBetan,2012prcIdBetan}.

The goal of this work is to complement Refs. \cite{1999VonDelft} and \cite{2012prcIdBetan} giving a guided demostration of how, the pair-energies and the Richardson's equations, emerge from the many-body eigenvalue equation.

In sections \ref{sec.2} and \ref{sec.3} we defined the model interaction and the solution for the many-body system, respectively. The complementary equations (Richardson equations) needed to find the many-body eigenenergy is given in section \ref{sec.4} for the box representation. The quasi-bound contiuum states in the box are changed by the single particle density in section \ref{sec.5} and the Richardson equations are given. Section \ref{sec.6} gives the Richardson equations for a complex energy representation. Some discusion are made in the last section \ref{sec.7}. Four appendixes complement this work by given some details for calculations.

\section{Pairing Hamiltonian}\label{sec.2}
The original derivation of the Richardson's equations in Refs. \cite{1963Richardson} and \cite{1964Richardson} is very cumbersome. From a Richardson's suggestion \cite{1999VonDelft}, Von Delft and Braun derived Richardson's equations in a simpler and clearer way. This last approach is adopted in this work to derived the equation with continuum spectrum and its extension to the complex energy plane. Due to the dificulties of dealing explicitly with the continuum, we will work out the equations by considering the system inmersed in an spherical box of dimension $R$. At the end of the calculations we will make the formal limit of the size of the box to infinity in order to have the final equations in the real energy representation. The final stage will consist to make analytic extension of the equations to the complex energy plane.

Let us assume that the single-particle levels of our representation has negative and positive energy $\varepsilon_j$, and that each state is doubly degenerate. The constant pairing Hamiltonian reads,
\begin{equation}
  H = \sum_{j \sigma} \varepsilon_j c^\dagger_{j \sigma} c_{j \sigma}
       - G \sum_{i j} c^\dagger_{j+} c^\dagger_{j-} c_{j-} c_{j+}\, ,
\end{equation}
with $|j,\pm \rangle$ time-reversed states, and $G$ the strength of the pairing interaction. The fermion operators $c^\dagger_{j \sigma}$ satisfy the usual anti-commutation relationship.

Now we introduce the pair creation operators $b^\dagger_j = c^\dagger_{j+} c^\dagger_{j-}$, which satisfy the following conmutation relations $[ b_{j},b^\dagger_{j'} ]=\delta_{jj'}(1-2 n_j)= \delta_{jj'}(1-2 b^\dagger_j b_j)$, with $n_j = c^\dagger_{j+} c_{j+}+c^\dagger_{j-} c_{j-}$. The last identity in the conmutation follows because, as it is explained in Ref. \cite{Schieffer} (pag. 38), '...the ground state of $H$ has no pair state $(j+,j-)$ occupied by a single fermion.' The other conmutations relations are nil. The operators $b^\dagger_j$ also satisfy $[ b^\dagger_{j}b_j,b^\dagger_{j'} ] = \delta_{jj'} b^\dagger_{j}$ and $( b^\dagger_j )^2 = 0$.

For practical porpuse needed in the next section we write the Hamiltonian in terms of the operators $B_0^\dagger = \sum_i b^\dagger_i$ and $B_0 = \sum_i b_i$, 
\begin{eqnarray}
  H &=& \sum_j 2 \varepsilon_j  b^\dagger_j b_j - G B_0^\dagger B_0
\end{eqnarray}

\section{Eigenvalue equation and many-body eigenenergy}\label{sec.3}
We start by making the following anzats for the $2n$-body wave function \cite{1999VonDelft}, where $n$ is the number of pairs in the system,
\begin{equation}\label{eq.wf}
  | \psi_n \rangle = \prod_{\nu=1}^n B_{J_\nu}^\dagger |0\rangle
  					=  \prod_{\nu=1}^n
  					\left(
  					 \sum_j \frac{b_j^\dagger}{2\varepsilon_j-E_{J_\nu}}
  					\right) |0\rangle
\end{equation}

Our goal is to find the many-body eigenenergy $\mathcal{E}_n$ and the coefficients $E_{J_\nu}$ which define the corresponding many-body eigenfunction. To reach this goal we solve the many-body Schr\"odinger equation with the above wave function
\begin{eqnarray}
  H | \psi_n \rangle &=& \mathcal{E}_n | \psi_n \rangle \, ,
\end{eqnarray}
which can be written in the following way
\begin{eqnarray}\label{eq.eiv2}
 \left[ H,\prod_{\nu=1}^n B_{J_\nu}^\dagger \right] |0\rangle &=& \mathcal{E}_n |\psi_n \rangle
\end{eqnarray}
where we have used $H |0\rangle=0$, i.e. the eigenvalues $\mathcal{E}_n$ of $H$ are defined with respect to the vacuum state $|0\rangle$.

Using the identity (\ref{eq.ident1}) of the \ref{sec.conm0} we write the commutator $[H,\prod_\nu B_{J_\nu}^\dagger]$ in terms of the single commutator $[H,B_J^\dagger]$
\begin{eqnarray}
 [ H,\prod_{\nu=1}^n B_{J_\nu}^\dagger ] &=&
   \sum_{\nu=1}^n
   \left\{
         \left( \prod_{\eta=1}^{\nu-1} B_{J_\eta}^\dagger \right)
         [ H, B_{J_\nu}^\dagger ]
         \left( \prod_{\mu=\nu+1}^{n} B_{J_\mu}^\dagger \right)
   \right\} 
\end{eqnarray}

By replacing the value of $[H,B_J^\dagger]$ (see Eq. \ref{eq.HBJ}), calculated in the \ref{sec.conm1} we get 
\begin{eqnarray}
 [ H,\prod_{\nu=1}^n B_{J_\nu}^\dagger ] &=&
   \sum_{\nu=1}^n
   \left\{
         \left( \prod_{\eta=1}^{\nu-1} B_{J_\eta}^\dagger \right)
         \left[
          E_{J_\nu} B^\dagger_{J_\nu} 
          + B_0^\dagger \left( 1 - G \sum_j \frac{1-2 b^\dagger_j b_j }{2\varepsilon_j-E_{J_\nu}} \right)
         \right]
         \left( \prod_{\mu=\nu+1}^{n} B_{J_\mu}^\dagger \right)
   \right\} \nonumber
\end{eqnarray}

By distributing the product and gathering the indexes when possible we get
\begin{eqnarray}
 [ H,\prod_{\nu=1}^n B_{J_\nu}^\dagger ] &=&
   \sum_{\nu=1}^n E_{J_\nu} \prod_{\eta=1}^n B_{J_\eta}^\dagger \nonumber \\
 &&+\sum_{\nu=1}^n \left[
                   1-\sum_j \frac{G}{2\varepsilon_j-E_{J_\nu}}
                   \right]
                   B_0^\dagger \prod_{\eta=1,\eta \ne \nu}^{n} B_{J_\eta}^\dagger \nonumber \\
 &&+\sum_{\nu=1}^n \left( \prod_{\eta=1}^{\nu-1} B_{J_\eta}^\dagger \right)
                   \sum_j \frac{2G B_0^\dagger b^\dagger_j b_j}{2\varepsilon_j-E_{J_\nu}}
                   \left( \prod_{\mu=\nu+1}^{n} B_{J_\mu}^\dagger \right)
\end{eqnarray}

Next we applied the above identity to the vacuum $|0 \rangle$ and we use  $\prod_{\eta=1}^n B_{J_\eta}^\dagger |0 \rangle = |\psi_n \rangle$ to get,
\begin{eqnarray}
 [ H,\prod_{\nu=1}^n B_{J_\nu}^\dagger ] |0 \rangle &=& 
   \sum_{\nu=1}^n E_{J_\nu} |\psi_n \rangle \nonumber \\
  &&+\sum_{\nu=1}^n \left[
                   1-\sum_j \frac{G}{2\varepsilon_j-E_{J_\nu}}
                   \right]
                   B_0^\dagger \prod_{\eta=1,\eta \ne \nu}^{n} B_{J_\eta}^\dagger |0 \rangle \nonumber \\
 &&+\sum_{\nu=1}^n \left( \prod_{\eta=1}^{\nu-1} B_{J_\eta}^\dagger \right)
                   \sum_j \frac{2G B_0^\dagger b^\dagger_j b_j}{2\varepsilon_j-E_{J_\nu}}
                   \left( \prod_{\mu=\nu+1}^{n} B_{J_\mu}^\dagger \right) |0 \rangle \nonumber \\
\end{eqnarray}

Using Eq. (\ref{eq.eiv2}) for the left hand side we have
\begin{eqnarray}\label{eq.long}
 \mathcal{E}_n |\psi_n \rangle &=&
   \sum_{\nu=1}^n E_{J_\nu} |\psi_n \rangle \nonumber \\
  &&+\sum_{\nu=1}^n \left[
                   1-\sum_j \frac{G}{2\varepsilon_j-E_{J_\nu}}
                   \right]
                   B_0^\dagger \prod_{\eta=1,\eta \ne \nu}^{n} B_{J_\eta}^\dagger |0 \rangle \nonumber \\
 &&+\sum_{\nu=1}^n \left( \prod_{\eta=1}^{\nu-1} B_{J_\eta}^\dagger \right)
                   \sum_j \frac{2G B_0^\dagger b^\dagger_j b_j}{2\varepsilon_j-E_{J_\nu}}
                   \left( \prod_{\mu=\nu+1}^{n} B_{J_\mu}^\dagger \right) |0 \rangle \nonumber \\
\end{eqnarray}

If the second and third lines of the above equation would be zero, the eigenenergy of the many-body system would be given in terms of the parameters $E_{J_\nu}$ by
\begin{equation}\label{eq.en}
  \mathcal{E}_n = \sum_{\nu=1}^n E_{J_\nu} 
\end{equation} 

In the next section we show which are the conditions that the parameters $E_{J_\nu}$ have to satisfy, in order that these 'dangerous' terms cancel out.

\section{Richardson equations}\label{sec.4}
In this section we are going to reduce the last two terms of the right hand of equation (\ref{eq.long}) to a more amenable expresion. As a consecuence, we are going to get the Richardson equations, which give the conditions that the parameters $E_{J_\nu}$ must to satisfy to validate Eq. (\ref{eq.en}).

Let us write the dangerous term as $R |0 \rangle = R_1 |0 \rangle + R_2 |0 \rangle$, with
\begin{eqnarray} \label{eq.R}
 R_1 &=&
  \sum_{\nu=1}^n \left[
                   1-\sum_j \frac{G}{2\varepsilon_j-E_{J_\nu}}
                   \right]
                   B_0^\dagger \prod_{\eta=1,\eta \ne \nu}^{n} B_{J_\eta}^\dagger  \\
 R_2 &=&
   \sum_{\nu=1}^n \left( \prod_{\eta=1}^{\nu-1} B_{J_\eta}^\dagger \right)
                   \sum_j \frac{2G B_0^\dagger b^\dagger_j b_j}{2\varepsilon_j-E_{J_\nu}}
                   \left( \prod_{\mu=\nu+1}^{n} B_{J_\mu}^\dagger \right) 
\end{eqnarray}

Next we are going to reduce $R_2 |0 \rangle$ to a structure similar to $R_1 |0 \rangle$. In the \ref{sec.conm2} we show that $\sum_j 2G B_0^\dagger b^\dagger_j b_j/(2\varepsilon_j-E_{J_\nu})(\prod_{\mu=\nu+1}^{n} B_{J_\mu}^\dagger)$ can be written as
\begin{eqnarray}
    \sum_{\nu'=\nu+1}^n
    \left\{
    \left( \prod_{\eta'=\nu+1}^{\nu'-1} B^\dagger_{J_{\eta'}} \right) \right. \nonumber 
  \sum_j \frac{2G B_0^\dagger b^\dagger_j}{(2\varepsilon_j-E_{J_\nu})(2\varepsilon_j-E_{J_{\nu'}})} 
 \left. \left( \prod_{\mu'=\nu'+1}^{n} B^\dagger_{J_{\mu'}} \right) \right\} 
\end{eqnarray}

Then
\begin{eqnarray}
   R_2 |0 \rangle &=& 
		\sum_{\nu=1}^n \left( \prod_{\eta=1}^{\nu-1} B_{J_\eta}^\dagger \right) 
  \sum_{\nu'=\nu+1}^n
    \left\{
    \left( \prod_{\eta'=\nu+1}^{\nu'-1} B^\dagger_{J_{\eta'}} \right) \right. 
 \sum_j \frac{2G B_0^\dagger b^\dagger_j}{(2\varepsilon_j-E_{J_\nu})(2\varepsilon_j-E_{J_{\nu'}})} 
 \left. \left( \prod_{\mu'=\nu'+1}^{n} B^\dagger_{J_{\mu'}} \right) \right\} |0 \rangle \nonumber
\end{eqnarray}

In the \ref{sec.conm3} we worked out the expression 
$\sum_j \frac{2G B_0^\dagger b^\dagger_j}{(2\varepsilon_j-E_{J_\nu})(2\varepsilon_j-E_{J_{\nu'}})}$ and got
\begin{eqnarray}
  \sum_j \frac{2G B_0^\dagger b^\dagger_j}{(2\varepsilon_j-E_{J_\nu})(2\varepsilon_j-E_{J_{\nu'}})}
  &=&
  \frac{2G B_0^\dagger}{E_{J_\nu} - E_{J_{\nu'}}} (B^\dagger_{J_{\nu}}-B^\dagger_{J_{\nu'}})
\end{eqnarray}

Then
\begin{eqnarray}
 R_2 |0 \rangle &=&
 		\sum_{\nu=1}^n \left( \prod_{\eta=1}^{\nu-1} B_{J_\eta}^\dagger \right) 
   \sum_{\nu'=\nu+1}^n
    \left\{
    \left( \prod_{\eta'=\nu+1}^{\nu'-1} B^\dagger_{J_{\eta'}} \right) \right. 
    \frac{2G B_0^\dagger}{E_{J_\nu} - E_{J_{\nu'}}} (B^\dagger_{J_{\nu}}-B^\dagger_{J_{\nu'}})
   \left. \left( \prod_{\mu'=\nu'+1}^{n} B^\dagger_{J_{\mu'}} \right) \right\} |0 \rangle \nonumber
\end{eqnarray}

This equation can be written as
\begin{eqnarray}
 R_2 |0 \rangle &=&
 	B_0^\dagger \sum_{\nu=1}^n \sum_{\nu'=\nu+1}^n
    \frac{2G}{E_{J_\nu} - E_{J_{\nu'}}}
    \prod_{\eta=1, \eta \ne \nu'}^{n} B_{J_\eta}^\dagger |0 \rangle \nonumber \\
 &&-B_0^\dagger \sum_{\nu=1}^n \sum_{\nu'=\nu+1}^n
    \frac{2G}{E_{J_\nu} - E_{J_{\nu'}}}
    \prod_{\eta=1, \eta \ne \nu}^{n} B_{J_\eta}^\dagger |0 \rangle
\end{eqnarray}

By changing the indexes of the summations in the first line using Eq. (\ref{eq.ident2}) and then changing  $\nu \leftrightarrow \nu'$ we get
\begin{eqnarray}
 R_2 |0 \rangle &=&
		B_0^\dagger \sum_{\nu=1}^n \sum_{\nu'=1}^{\nu-1}
    \frac{2G}{E_{J_{\nu'}} - E_{J_{\nu}}}
    \prod_{\eta=1, \eta \ne \nu}^{n} B_{J_\eta}^\dagger |0 \rangle \nonumber \\
 &&-B_0^\dagger \sum_{\nu=1}^n \sum_{\nu'=\nu+1}^n
    \frac{2G}{E_{J_\nu} - E_{J_{\nu'}}}
    \prod_{\eta=1, \eta \ne \nu}^{n} B_{J_\eta}^\dagger |0 \rangle
\end{eqnarray}

Finally, by inverting the term $E_{J_{\nu'}} - E_{J_{\nu}}$ in the first line and by merging the summation in $\nu'$ we get
\begin{eqnarray}
 R |0 \rangle &=&
   \sum_{\nu=1}^n
    \left\{
    \left[
     1
     -\sum_j \frac{G}{2\varepsilon_j-E_{J_\nu}}
     -\sum_{\nu'=1, \nu' \ne \nu}^{n} \frac{2G}{E_{J_{\nu}} - E_{J_{\nu'}}}
    \right]
    B_0^\dagger \prod_{\eta=1,\eta \ne \nu}^{n} B_{J_\eta}^\dagger
    \right\} |0 \rangle \nonumber \\
\end{eqnarray}

Then, the conditions which have to satisfy the parameters $E_{J_\nu}$ with $\nu=1,\cdots,n$ (and $n$ the number of pairs), is that all $E_{J_\nu}$ simultaneously satisfy the $n$ non-linear equations, i.e. the Richardson equations \cite{1963Richardson}
\begin{equation}\label{eq.rich}
  1 - \sum_j \frac{G}{2\varepsilon_j-E_{J_\nu}}
  -\sum_{\nu'=1, \nu' \ne \nu}^{n} \frac{2G}{E_{J_{\nu}} - E_{J_{\nu'}}} = 0
\end{equation}

This complete the exact solution of the many-body pairing Hamiltonian with continuum in a box representation. In the next section we will make the limit to the continuum.

\section{Real continuum energy spectrum}\label{sec.5}
In this section we make the formal limit of the size $R$ of the spherical box to infinity. In doing so, the negative energy spectrum stabilizes while the continuum spectrum became denser as the size of the box increases. For the infity limit we make the following substitution
\begin{equation}
 \sum_{j \sigma} \xrightarrow{R\rightarrow \infty} 
          \int_{-\infty}^\infty \; \tilde{g}(\varepsilon) \; d\varepsilon \;.
\end{equation}
where the single particle density $\tilde{g}(\varepsilon)$ has two terms, (i) the first due to negative energy $\varepsilon_{j_b}$ of bound states, and (ii) the second due to the continuum states $\varepsilon$ with positive energy and density $g(\varepsilon)$,
\begin{equation}
    \tilde{g}(\varepsilon) = \sum_{j_b \sigma} \; \delta(\varepsilon-\varepsilon_{j_b}) +
                         g(\varepsilon)
\end{equation}

The Richardson equations in a representation with real continuum energy (besides the bound states) reads
\begin{eqnarray} \label{eq.rich_cont}
 1 - \frac{G}{2} \sum_{j_b \sigma} \frac{1}{2 \varepsilon_b - E_{\nu}} 
   - \frac{G}{2} \int_0^\infty d\varepsilon  \; \frac{g(\varepsilon)}{2 \varepsilon - E_{\nu}}  
    + 2 G \sum_{\nu \ne \nu'} \; \frac{1}{E_{\nu} - E_{{\nu'}}} = 0 \;.
\end{eqnarray}
note that unlike Eq. (\ref{eq.rich}), here the degeneracy was included in the sum symbol. There are as many Eq. (\ref{eq.rich_cont}) as pairs in the system; all these equations have to be satisfied simultaneously in order to get all pair-energies. Then, summing up the $n$ pair-energies we get eigenenergy for the $2n$-body system through Eq. (\ref{eq.en}). This result also implies that only the formulas which define the Richardson equations change in a continuum representation but not the formula which gives the eigenenergy.

\section{Complex continuum energy spectrum}\label{sec.6}
In this section we separate from the density, the resonant part from the non-resonant part. It is expected that if the system holds narrow resonances, them will be the most important correlations coming from the continuum configurations. For the single particle density we use the one given by Beth and Uhlenbeck \cite{1937Beth} in terms of the phase shift,
\begin{equation}
   g(\varepsilon) = \sum_{j_c \sigma} \; \frac{1}{\pi} \; \frac{d\delta_c}{d\varepsilon}
      = g_{_{\rm Res}}(\varepsilon) + g_{_{\rm Bckg}}(\varepsilon) 
\end{equation}

The presence of single particle resonances will show up in the density, and also in the cross section, as sharp structures. These sharp structures may be parametrized using the Lorentzian distribution in terms of the resonant energy $\epsilon_r$ and the resonant width $\Gamma_r$
 \cite{1988Kukulin2}
\begin{equation}
 g_{_{\rm Res}}(\varepsilon) \approx \sum_{j_r \sigma} \frac{1}{\pi} \frac{\Gamma_r/2}{(\varepsilon - \epsilon_r)^2+(\Gamma_r/2)^2} \;.
\end{equation}
The above approximation worsen as the width of the resonance increases.

The analytic extension of the resonant density $g_{_{\rm Res}}(\varepsilon)$ to the lower complex energy plane, shows a pole  at the complex energy $\varepsilon_r = \epsilon_r - i \; \Gamma_r/2$. By rotating the integration contour of the  resonant part of the density in the Richardson equation (\ref{eq.rich_cont}) we get a contribution similar to the bound states but with complex energies instead. The following expression gives the Richardson equations in the complex energy representation \cite{2012prcIdBetan}
\begin{eqnarray}\label{eq.rich_cmpx}
  1 - \frac{G}{2} \sum_{j_b \sigma} \frac{1}{2 \varepsilon_b - E_{\nu}} 
   - \frac{G}{2} \sum_{j_r \sigma} \frac{1}{2 \varepsilon_r - E_{\nu}} 
   - \frac{G}{2} \int_0^\infty d\varepsilon  \; \frac{g_{_{\rm Bckg}}(\varepsilon)}{2 \varepsilon - E_{\nu}} 
   - \frac{G}{2} \int_0^\infty d\varepsilon  \; \frac{g_{_{\rm CxBckg}}(\varepsilon)}{2 \varepsilon - iE_{\nu}} \nonumber \\
     + 2 G \sum_{\nu \ne \nu'} \; \frac{1}{E_{\nu} - E_{\nu'}} = 0 \;,
\end{eqnarray}
where $g_{_{\rm CxBckg}}(\varepsilon)$ is the density which remains after the poles contribution of the resonant part of the density was taken.

The solution of the Richardson equations (\ref{eq.rich_cmpx}) give the pair-energies to build the many-body eigenenergy from Eq. (\ref{eq.en}). If we ignore the densities $g_{_{\rm Bckg}}$ and $g_{_{\rm CxBckg}}$ in Eq. (\ref{eq.rich_cmpx}), we end up with the pole approximation of the Richardson equations given in Ref. \cite{2003Hasegawa}. Note that in this approximation, the eigenenergy given by Eq. (\ref{eq.en}) will no be any more real. The magnitude of the imaginary component of $\mathcal{E}_n$ is a measure of how good (or bad) the pole approximation is.

\section{Conclusions and discussion}\label{sec.7}
We have shown how to  get the many-body eigenvalue for the constant pairing Hamiltonian for different model spaces including the continuum spectrum of energy. The procedure followed consisted of solving the many-body Schr\"odinger equation with the ansatz wave function (\ref{eq.wf}) of Ref. \cite{1996VonDelft}. The equations obtained are exact in the box representation, i.e., before the limit of the radius of the spherical box is taken to infinity. In the limit, a single particle density has to be introduced an some arbitrariness is given to the model solution depending of the density used. Since the total density diverges with the same divergence as the free density as the radius of the spherical box goes to infinity \cite{1997Shlomo}, we decided to take as density the difference between the mean-field and the free densities \cite{1937Beth}, i.e., in terms of the derivatives of the phase shift. Another advantage of this density is that it makes very natural the separation between resonan continuum and nonresonant continuum, which allow to use a simple parametrization for the resonant configurations. As a last step, in order to move from the real continuum to the complex continuum, we made the analytic continuation of the parametrized density and by using Complex Analysis we were able to changed the integral contribution of the resonant continuum by a single term (for each partial wave).

\ack
This work has been supported by the Consejo Nacional de Investigaciones Cient\'{\i}ficas y T\'ecnicas PIP-625 (CONICET, Argentina).

\appendix
\section{Math tool box}\label{sec.conm0}
This appendix collect three identities which are used in the process of derivation of the Richardson equations.

\paragraph{Isolation of one commutator:} The following expression give the identity to factorize a single commutator from a commutator involving product of operators \cite{1999VonDelft}
\begin{equation}\label{eq.ident1}
  [A,\prod_{i=1}^n B_i] = \sum_{i=1}^n 
                      \left\{
                       \left(\prod_{j=1}^{i-1} B_j\right)
                       [A,B_i]
                       \left(\prod_{j'=i+1}^{n} B_{j'}\right)
                      \right\}
\end{equation}

\paragraph{Changing indexes in double sums:} By visualizing a double sum as the sum of the matrix elements of a matrix of order $n$, follows the identity below for the sum of all non diagonal terms
\begin{equation}\label{eq.ident2}
  \sum_{i=1}^n \sum_{j=i+1}^n a_{ij} =  \sum_{j=1}^n \sum_{i=1}^{j-1} a_{ij}
\end{equation}

\paragraph{Partial fraction expansion:} The partial fraction decomposition allows to write the quotient of two polynomials as a sum of terms with simpler numerator,
\begin{eqnarray}\label{eq.ident3}
  \frac{P(x)}{Q(x)} &=& \frac{P(x)}{\prod_i (x-x_i)} 
                    = \sum_i \frac{c_i}{x-x_i}
\end{eqnarray}
where $x_i$ are the zeros of $Q(x)$. In order to calculate the coefficients $c_i$ one has to evaluate some expression at the values of the roots $x_i$.

\section{Calculation of $[H,B_J^\dagger]$}\label{sec.conm1}
\begin{eqnarray}
  [H,B_J^\dagger] &=& \sum_j 2 \varepsilon_j [b^\dagger_j b_j, B_J^\dagger]
                      - G [B_0^\dagger B_0,B_J^\dagger]
\end{eqnarray}

The commutation in first term reduces to
\begin{eqnarray}
  [b^\dagger_j b_j, B_J^\dagger] &=& b^\dagger_j [b_j, B_J^\dagger]
                                     -[b^\dagger_j, B_J^\dagger] b_j \\
     &=& b^\dagger_j \sum_{j'} \frac{1}{2\varepsilon_{j'}-E_J} [b_j, b_{j'}^\dagger]
        - \sum_{j'} \frac{1}{2\varepsilon_{j'}-E_J} [b^\dagger_j, b_{j'}^\dagger] b_j \\
     &=& b^\dagger_j \sum_{j'} \delta_{jj'} \frac{1-2b^\dagger_j b_j}{2\varepsilon_{j'}-E_J}
        - \sum_{j'} \frac{1}{2\varepsilon_{j'}-E_J} 0 b_j 
     = \frac{b^\dagger_j-2 (b^\dagger_j)^2 b_j}{2\varepsilon_{j'}-E_J} \\
   &=& \frac{b^\dagger_j}{2\varepsilon_j-E_J}
\end{eqnarray}

While for the commutation of the second term we get
\begin{eqnarray}
  [B_0^\dagger B_0,B_J^\dagger] &=&
           B_0^\dagger [B_0,B_J^\dagger]
           - [B_0^\dagger ,B_J^\dagger]B_0
\end{eqnarray}
with
\begin{eqnarray}
  [B_0,B_J^\dagger] &=&
       [ \sum_j b_j , \sum_{j'} \frac{b^\dagger_{j'}}{2\varepsilon_{j'}-E_J} ] 
   = \sum_{jj'} \frac{1}{2\varepsilon_{j'}-E_J} [ b_j , b^\dagger_{j'} ] \\
   &=&  \sum_{j} \frac{1-2 b^\dagger_j b_j}{2\varepsilon_j-E_J}
\end{eqnarray}
and 
\begin{eqnarray}
  [B_0^\dagger,B_J^\dagger] &=&
       [ \sum_j b^\dagger_j , \sum_{j'} \frac{b^\dagger_{j'}}{2\varepsilon_{j'}-E_J} ] 
    = \sum_{jj'} \frac{1}{2\varepsilon_{j'}-E_J} [ b^\dagger_j , b^\dagger_{j'} ] \\
   &=& 0
\end{eqnarray}

Using the above identities, the commutator $[H,B_J^\dagger]$ reads
\begin{eqnarray}
  [H,B_J^\dagger] &=& 
          \sum_j \frac{2\varepsilon_j b^\dagger_j }{2\varepsilon_j-E_J}
          - G B^\dagger_0 \frac{1-2 b^\dagger_j b_j }{2\varepsilon_j-E_J}
\end{eqnarray}

The term $\sum_j 2\varepsilon_j b^\dagger_j/(2\varepsilon_j-E_J)$ can be written in terms of $B^\dagger_J$and $B_0^\dagger$,
\begin{eqnarray}
  \sum_j \frac{2\varepsilon_j b^\dagger_j }{2\varepsilon_j-E_J} &=&
          \sum_j \frac{2\varepsilon_j b^\dagger_j }{2\varepsilon_j-E_J} - B_0^\dagger + B_0^\dagger \\
   &=& \sum_j b^\dagger_j \left[ \frac{2\varepsilon_j }{2\varepsilon_j-E_J} - 1 \right] 
               + B_0^\dagger \\
   &=& E_J \sum_j b^\dagger_j \frac{1}{2\varepsilon_j-E_J} + B_0^\dagger \\
   &=&
      E_J B^\dagger_J + B_0^\dagger 
\end{eqnarray}

Gathering all together we get
\begin{eqnarray}\label{eq.HBJ}
  [H,B_J^\dagger] &=& 
   E_J B^\dagger_J 
        + B_0^\dagger \left( 1 - G \sum_j \frac{1-2 b^\dagger_j b_j }{2\varepsilon_j-E_J} \right) \, .
\end{eqnarray}

\section{Calculation of the commutator 
$[\sum_j 2G B_0^\dagger b^\dagger_j b_j/(2\varepsilon_j-E_{J_\nu}),\prod_\mu B_{J_\mu}^\dagger]$}\label{sec.conm2}
In order to invert the order of $b^\dagger_j b_j$ and $B_{J_\mu}^\dagger$ in eq. (\ref{eq.R}) we need to work out the following equation
\begin{eqnarray}
 [ \sum_j \frac{2G B_0^\dagger b^\dagger_j b_j}{2\varepsilon_j-E_{J_\nu}},
   \prod_{\mu=\nu+1}^{n} B_{J_\mu}^\dagger ] |0 \rangle &=& 
 \sum_j \frac{2G B_0^\dagger b^\dagger_j b_j}{2\varepsilon_j-E_{J_\nu}}
    \left( \prod_{\mu=\nu+1}^{n} B_{J_\mu}^\dagger \right) |0 \rangle
\end{eqnarray}

By applying the indentity (\ref{eq.ident1}) we get
\begin{eqnarray}
 [ \sum_j \frac{2G B_0^\dagger b^\dagger_j b_j}{2\varepsilon_j-E_{J_\nu}},
                    \prod_{\mu=\nu+1}^{n} B_{J_\mu}^\dagger ] |0 \rangle &=& 
   \sum_{\nu'=\nu+1}^n
   \left\{
   \left( \prod_{\eta'=\nu+1}^{\nu'-1} B^\dagger_{J_{\eta'}} \right) \right. 
  [ \sum_j \frac{2G B_0^\dagger b^\dagger_j b_j}{2\varepsilon_j-E_{J_\nu}},B^\dagger_{J_{\nu'}}] \nonumber \\
 &&\left. \left( \prod_{\mu'=\nu'+1}^{n} B^\dagger_{J_{\mu'}} \right) \right\} |0 \rangle
\end{eqnarray}

where
\begin{eqnarray}
  [ \sum_j \frac{2G B_0^\dagger b^\dagger_j b_j}{2\varepsilon_j-E_{J_\nu}},B^\dagger_{J_{\nu'}}] |0 \rangle &=&
 \sum_j \frac{2G}{2\varepsilon_j-E_{J_\nu}}
 \left\{ B_0^\dagger [b^\dagger_j b_j,B^\dagger_{J_{\nu'}}] \right. \nonumber \\
 && - \left. [ B_0^\dagger, B^\dagger_{\nu'}] b^\dagger_j b_j \right\} |0 \rangle \\
 &=&
  \sum_j \frac{2G}{2\varepsilon_j-E_{J_\nu}} B_0^\dagger
         \frac{b^\dagger_j}{2\varepsilon_j-E_{J_{\nu'}}} |0 \rangle \, .
\end{eqnarray}

Then,
\begin{eqnarray}
 \sum_j \frac{2G B_0^\dagger b^\dagger_j b_j}{2\varepsilon_j-E_{J_\nu}}
    \left( \prod_{\mu=\nu+1}^{n} B_{J_\mu}^\dagger \right) |0 \rangle =
   \sum_{\nu'=\nu+1}^n
   \left\{
   \left( \prod_{\eta'=\nu+1}^{\nu'-1} B^\dagger_{J_{\eta'}} \right) \right. 
  \sum_j \frac{2G B_0^\dagger b^\dagger_j}
                {(2\varepsilon_j-E_{J_\nu})(2\varepsilon_j-E_{J_{\nu'}})} \nonumber \\
 \left. \left( \prod_{\mu'=\nu'+1}^{n} B^\dagger_{J_{\mu'}} \right) \right\} |0 \rangle
\end{eqnarray}

\section{Calculation of the term 
$\sum_j \frac{2g B_0^\dagger b^\dagger_j}{(2\varepsilon_j-E_{J_\nu})(2\varepsilon_j-E_{J_{\nu'}})}$}\label{sec.conm3}
By using the simple fraction expansion of Eq. (\ref{eq.ident3}) we write,
\begin{eqnarray}
  \sum_j \frac{2g B_0^\dagger b^\dagger_j}{(2\varepsilon_j-E_{J_\nu})(2\varepsilon_j-E_{J_{\mu}})} 
    &=&
    2g B_0^\dagger \sum_j b^\dagger_j \frac{1}{(2\varepsilon_j-E_{J_\nu})(2\varepsilon_j-E_{J_{\mu}})} \nonumber \\
 &=& 2g B_0^\dagger \sum_j b^\dagger_j 
    \left[
     \frac{A}{2\varepsilon_j-E_{J_\nu}}
     + \frac{B}{2\varepsilon_j-E_{J_\mu}}
    \right] \nonumber \\
 &=& 2g B_0^\dagger \sum_j b^\dagger_j 
    \left[
     \frac{A(2\varepsilon_j-E_{J_\mu})+B(2\varepsilon_j-E_{J_\nu})}
          {(2\varepsilon_j-E_{J_\nu})(2\varepsilon_j-E_{J_\nu})}
    \right] \nonumber
\end{eqnarray}
with $1=A(2\varepsilon_j-E_{J_\mu})+B(2\varepsilon_j-E_{J_\nu})$. For $2\varepsilon_j=E_{J_\mu}$ we get $B= \frac{1}{E_{J_\mu}-E_{J_\nu}}$, while for $2\varepsilon_j=E_{J_\nu}$ we get $A= \frac{1}{E_{J_\nu}-E_{J_\mu}} = -B$

Then
\begin{eqnarray}
  \sum_j \frac{2g B_0^\dagger b^\dagger_j}{(2\varepsilon_j-E_{J_\nu})(2\varepsilon_j-E_{J_{\mu}})} 
 &=& 2g B_0^\dagger
    \sum_j b^\dagger_j
    \frac{1}{E_{J_\nu}-E_{J_\mu}}
    \left(
    \frac{1}{2\varepsilon_j-E_{J_\nu}}
    -\frac{1}{2\varepsilon_j-E_{J_\mu}}
    \right) \nonumber \\
 &=& \frac{2g B_0^\dagger}{E_{J_\nu}-E_{J_\mu}}
     \left[
     \sum_j \frac{b^\dagger_j}{2\varepsilon_j-E_{J_\nu}}
     -\sum_j \frac{b^\dagger_j}{2\varepsilon_j-E_{J_\mu}}
    \right] \nonumber \\
 &=& \frac{2g B_0^\dagger}{E_{J_\nu}-E_{J_\mu}}
     \left( B^\dagger_{J_\nu} - B^\dagger_{J_\mu} \right) \, .
\end{eqnarray}

\section*{References}
\providecommand{\newblock}{}

\end{document}